\documentclass[%
reprint,
superscriptaddress,
nofootinbib,
amsmath,
amssymb,
aps,
floatfix,
showkeys,
]{revtex4-2}

\usepackage[colorlinks,citecolor=blue,linkcolor=blue,anchorcolor=blue,filecolor=blue,
urlcolor=blue]{hyperref}

\usepackage{graphicx,color,xcolor}
\usepackage{amsfonts,amsmath}
\usepackage{amssymb,ulem}
\usepackage{dcolumn}
\usepackage{bm,lipsum}
\usepackage[utf8]{inputenc}
\usepackage{subeqnarray}
\usepackage{array}
\usepackage{epsfig,hyperref}
\usepackage{morefloats}
\usepackage{relsize}
\usepackage{url}
\usepackage{orcidlink}
\usepackage{lineno}

\usepackage{comment}

\begin{document}

\preprint{ }

\title{ Oscillatory properties of strange quark stars described by the vector MIT bag model}

\author{Luiz L. Lopes
 }
\email{llopes@cefetmg.br}

\affiliation{Centro Federal de Educação Tecnológica de Minas Gerais Campus VIII, \\ CEP 37.022-560, Varginha, MG, Brazil}

\author{ José C. Jiménez 
}
\affiliation{Departament of Astrophysics, Brazilian Center for Research in Physics (CBPF),\\ Rua Dr. Xavier Sigaud, 150, URCA, Rio de Janeiro CEP 22210-180, RJ, Brazil}
\affiliation{Universidad Tecnológica del Perú, Arequipa - Perú}

\author{Luis B. Castro   
}
\email{llopes@cefetmg.br}
\affiliation{Universidade Federal do Maranhão, UFMA, Departamento de Física-CCET, Campus Universitário do Bacanga, São Luís, CEP 65080-805 Maranhão, Brazil}

\author{César V. Flores 
}
\email{cesar.vasquez@uemasul.edu.br}
\affiliation{Universidade Estadual da Região Tocantina do Maranhão, UEMASUL, Centro de Ciências Exatas, Naturais e Tecnológicas, Imperatriz, CEP 65901-480 Maranhão, Brazil}
\affiliation{Universidade Federal do Maranhão, UFMA, Departamento de Física-CCET, Campus Universitário do Bacanga, São Luís, CEP 65080-805 Maranhão, Brazil}

\begin{abstract}
We investigated the radial and non-radial fundamental ($f$) mode oscillations of self-bound (quark) stars obtained after employing the Vector MIT (vMIT) bag model. Within this model, we computed the equation of state for strange quark matter satisfying thermodynamic consistency. This allowed us to obtain the corresponding behavior of the speed of sound, mass-radius relation, and gravitational redshift. In particular, our choice of $G_V$ = 0.30 fm$^2$ produces masses and radii in agreement with recent astronomical data (e.g. from NICER and HESS J1731). In fact, we tested that variations of the remaining vMIT parameters slightly modify this conclusion. Then, we proceeded to compute the radial oscillation frequencies of the $f$-mode, which is tightly connected to the dynamical stability of these compact stars. We found that increments of the $G_V$ parameter have a stabilizing property around the maximal-mass stars for a given stellar family. We also calculated the gravitational-wave frequencies of the non-radial $f$-mode. Our results show that they are restricted to be in the range (1.6 - 1.8) kHz for high-mass stars and to (1.5 - 1.6) kHz for low-mass stars. Finally, we propose a universal relation between these frequencies and the square root of the average density. All these last results are important in distinguishing strange stars from ordinary neutron stars in future gravitational-wave detections coming from compact sources with activated non-radial modes.

\end{abstract}

\maketitle


\section{Introduction} \label{introduction}

Strange quark stars are self-bounded compact objects composed of deconfined quarks at their interiors. The theory of strange quark stars is based on the so-called Bodmer–Witten conjecture~\cite{Bodmer,Witten}, which assumes that ordinary matter, composed of protons and neutrons, may only be meta-stable phase, while the true ground state of strongly interacting matter would therefore consist
of the so-called strange quark matter (SQM), i.e. matter constituted by deconfined up, down, and strange quarks. 

Assuming that at least some of the observed pulsars can potentially be strange quark stars, in this work, we study some macroscopic and microscopic properties of these objects. Using the vMIT bag model~\cite{lopesps1,lopesps2}, we begin by studying the equation of state (EoS) and its corresponding speed of sound for different values of the (main) coupling $G_V$. Afterwards, using the EoS as input, we study their macroscopic observables such as the mass-radius relation and the gravitational redshift. We then probe if these SQM stars fulfill current observational constraints, e.g. for their maximal masses \cite{NANOGrav:2019jur}.

In addition to the macroscopic relations that can be studied via electromagnetic signals, e.g. X-ray data for the radii of these objects, additional information can be obtained from gravitational wave asteroseismology. A gravitational wave (GW) signal is a ripple in the fabric of space-time being produced by the accelerated motion of massive astrophysical objects, e.g. binary systems composed of two neutron stars. In fact, GWs were already detected in an indirect form near the year 1970 by observing the period of pulsar PSR B1913+16. Interestingly, in September 2015, a direct detection was performed by the LIGO-Virgo collaboration. Since then, hundreds of events have been observed including the merging of binary neutron stars \cite{LIGOScientific:2017vwq}. In 2019, the LIGO-Virgo-KAGRA collaboration was established and their observation run O4 is currently ongoing, with another run (O5) planned to be completed by the end of this decade. Last but not least, the next decade will mark the advent of third-generation terrestrial gravitational-wave observatories (Einstein Telescope \cite{Branchesi:2023mws} and Cosmic Explorer \cite{Evans:2021gyd}), as well as space detectors like LISA, opening the milliHertz frequency window \cite{Colpi:2024xhw}. The scientific knowledge expected from such intense research is outstanding, from fundamental physics to astrophysics and cosmology.

One of the phenomena that can produce GWs is the non-radial oscillations of compact stars. These oscillations can be categorized into different modes: \textit{f}-modes, \textit{p}-modes (pressure modes), and \textit{g}-modes (gravity modes). In this study, we focus our study on the \textit{f}-modes, whose frequency and damping time are directly influenced by the EoS of dense matter. Therefore, these modes directly describe how matter behaves at extremely high densities found in the inner regions of neutron stars. In this sense, it is well known that one can gain valuable insights of the microphysics and composition of neutron stars by analyzing gravitational waves generated by $f$-modes, since they are more easily excited compared to the overtones \cite{Flores2020}.

In this work, Sec. \ref{sec_MIT} outlines the use of the vMIT bag model to compute the EoS for SQM. Section \ref{secIII} then addresses the calculation of equilibrium configurations for each SQM star, followed by the presentation of mass-radius relationship curves for all our stellar models of self-bound matter. In Sec. \ref{secIV}, we compute the fundamental radial oscillation mode to ensure our study is restricted to dynamically stable stars. In Sec. \ref{secV}, we give a universal relation connecting focus, the quadrupolar \textit{f}-mode and the average density at the SQM star interiors. This is relevant since it is the most studied mode in the field of gravitational wave astronomy, thus being a promising source of gravitational radiation. Finally, Sec. \ref{conclusions} presents our summary and conclusions.

\section{The Vector MIT bag model} \label{sec_MIT}

The vector MIT bag model is an extension of the original MIT bag model~\cite{mitbag} that incorporates some
features of the quantum Hadrodynamics (QHD)~\cite{Serot_1992}. In its original form, the MIT bag model considers that each baryon is composed of three non-interacting quarks inside a bag. The
bag, in turn, corresponds to an infinite potential that confines the quarks. As a consequence, the quarks are free inside the bag and are forbidden to reach its exterior. All the information about the strong force relies on the bag pressure value, which mimics the vacuum pressure.

In the vector MIT bag model, the quarks are still confined inside the bag, but now they interact with each other through a { generic massive vector field. This vector field plays an analog role to the $\omega$ meson of the QHD~\cite{Serot_1992}.} Moreover, if one desires, the contribution of the Dirac sea can be taken into account through a self-interaction of the vector field~\cite{furnstahl1997vacuum}, although we do not follow this path in the present work, once its presence softens the EoS. Our Lagrangian of the vector MIT bag model, therefore, consists of the Lagrangian of the original MIT, plus the Yukawa-type Lagrangian of the vector field exchange. We must also add a mass term to maintain the thermodynamic consistency. It then reads~\cite{lopesps1,lopesps2}:

\begin{equation}
    \mathcal{L} =  \mathcal{L}_{\rm MIT} +  \mathcal{L}_V,  \label{l1}
\end{equation}
where
\begin{equation}
\mathcal{L}_{\rm MIT} = \sum_{i}\{ \bar{\psi}_i  [ i\gamma^{\mu} \partial_\mu - m_i ]\psi_i - B \}\Theta(\bar{\psi_i}\psi_i), \label{e1}
\end{equation}  

\begin{equation}
 \mathcal{L}_V = \sum_{i}\{\bar{\psi}_i g_{iV}(\gamma^\mu V_{\mu})\psi_i - \frac{1}{2}m_V^2 V^\mu V_\mu \} \Theta(\bar{\psi}_i\psi_i) ,\label{Ne9}
\end{equation}
{ where $\psi_i$ is the Dirac quark field, $B$ is the constant vacuum pressure, $m_V$ is the mass of the $V_\mu$ field, $g_{iV}$ is the coupling constant of the quark $i$ with the field $V_\mu$. The  $\Theta(\bar{\psi}_i\psi_i)$ is the Heaviside step function included to assure that the quarks exist only confined to the bag.  It is important to emphasize that although the vector field $V^u$ plays the same role as the $\omega$ meson in the QHD, this analogy is only formal. Unlike the $\omega$ meson, the $V^\mu$ field does not necessarily correspond to any real meson field.}

Applying the mean-field approximation (MFA)~\cite{Serot_1992}, and the Euler-Lagrange equations, we can obtain the energy eigenvalue for the  quark fields, and the equation of motion for the field $V_0 = \langle V_\mu \rangle \delta_0^\mu$ field:

\begin{equation}
 E_i =  \sqrt{m_i^2 + k_i^2} +g_{iV}V_0 \label{eigen}   
\end{equation}

\begin{equation}
m_V^2V_0 = \sum_i g_{iV}n_i\label{V0} ,
\end{equation}

\begin{equation}
n_i = \gamma_i \frac{k_f^3}{3\pi^2}  
\end{equation}
where $n_i$ is the number density of the $i$-th quark and $\gamma_i$ = 6 = $ (3\times 2)$ due to the number of colors and spin projections.
Moreover, at $T = 0$, the energy eigenvalue is also the chemical potential ($\mu$).
To construct an EoS in the MFA,  we now consider the Fermi-Dirac distribution of the quarks, and the Hamiltonian for the vector field and the bag pressure value, $\mathcal{H} = -\langle \mathcal{L} \rangle$. We obtain:

\begin{equation}
\varepsilon_i = \frac{\gamma_i}{2\pi^2}\int_0^{k_f} E_i ~k^2 dk,\label{nl4}
\end{equation}

\begin{equation}
\varepsilon =  \sum_i \varepsilon_i + B - \frac{1}{2}m_V^2V_0^2\,. \label{nl5}
\end{equation}
\noindent The last term of equation (\ref{nl5}) being absent in Refs.\cite{MNRAS485:4873:2019,APJ877:139:2019,MNRAS463:571:2016} is crucial to kept the thermodynamic consistency of the model. To construct an electrically neutral, beta-stable matter, leptons are added as a free Fermi gas. The pressure is obtained via the thermodynamic relation, $p = \sum_i \mu_i n_i - \varepsilon$, where the sum runs over all the fermions.

In the following, we redefine $G_V~\equiv~(g_{uV}/m_V)^2$ and define $X_V~\equiv~(g_{sV}/g_{uV})$. Here, we use a universal coupling for $X_V$,  $X_V$ = 1.0, in opposition to $X_V$ = 0.4 predicted by the symmetry group and used in Ref~\cite{Fran2024PRD,Carline2023BJP}. A universal coupling produces more massive stars for the same value of $G_V$. The mass values utilized in this work are the same as suggested in Refs.~\cite{lopesps1,lopesps2}; $m_{u}=m_{d}=4$\,MeV and $m_{s}=95$\,MeV.

Now, the values of $G_V$ and the bag are not fully independent. For the SM hypothesis to be true, the energy per baryon of the deconfined phase (for p = 0 and T = 0) must be lower than the nonstrange infinite baryonic matter. Or explicitly~\citep{Bodmer,Witten}:

\begin{equation}
 E_{(uds)}/A < 930 ~ \mbox{MeV}, \label{EL20} 
\end{equation}
at the same time, the nonstrange matter still needs to have  an energy per baryon higher than the one of nonstrange infinite baryonic matter, otherwise, protons and neutrons would decay into $u$ and $d$ quarks:
\begin{equation}
 E_{(ud)}/A > 930 ~\mbox{MeV} \label{EL21} .
\end{equation}

Therefore, both, eq.~(\ref{EL20}) and (\ref{EL21}) must  simultaneous
be true. For a given value of $G_V$, the values of the bag that satisfy the SQM hypothesis form the so-called stability window. The stability window for $G_V = 0.0$ up to 0.3 fm$^2$ is presented as Fig. 4 in Ref.~\cite{lopesps1}. The $G_V$ and the bag act together in a positive feedback way: If we increase the $G_V$ we increase the maximum mass of the star but also reduce the value of the bag inside the stability window. At the same time, lower values of the bag produce an additional increase in the maximum mass.

We use here three different values of $G_{V}$, $0.18$, $0.24$, and $0.30$ fm$^2$, and in order to satisfy the SQM hypothesis we see that the corresponding values for the Bag constant are: $B^{1/4}$ = 150, 145 and 140 MeV, respectively. Therefore, all these values are in agreement with the stability window; satisfy some observational constraints for the mass and radius of observed compact stars, as discussed below, and are in agreement with the Bayesian analysis presented in Ref.~\cite{Fran2024PRD}. Also, it is important to explain that the values of the bag constant are very near each other, so we easily see that the effect on the EoS, stellar masses, and gravitational frequencies are, mainly, due to the $G_V$ parameter.

\begin{figure}[ht]
  \begin{centering}
\begin{tabular}{c}
\includegraphics[width=0.333\textwidth,angle=270]{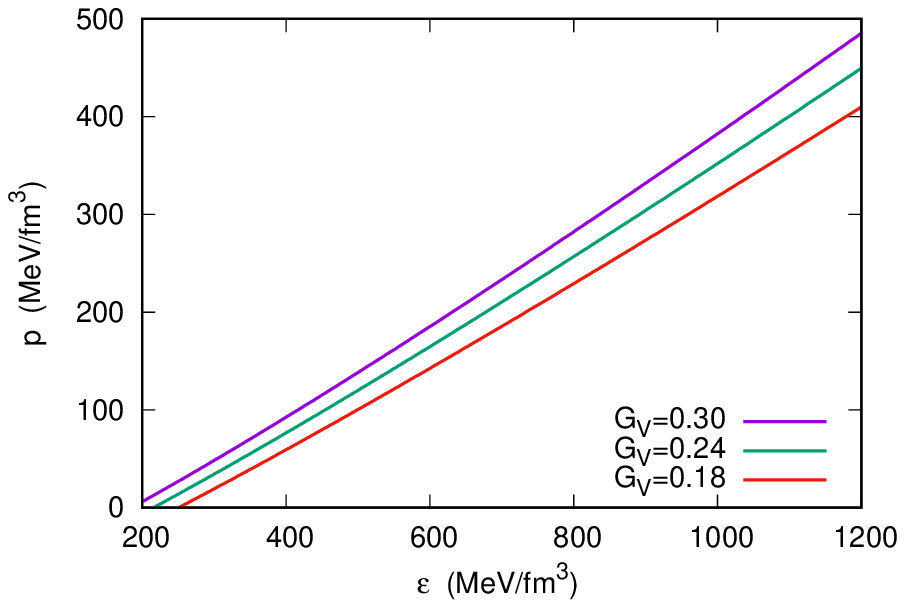} \\
\includegraphics[width=0.333\textwidth,,angle=270]{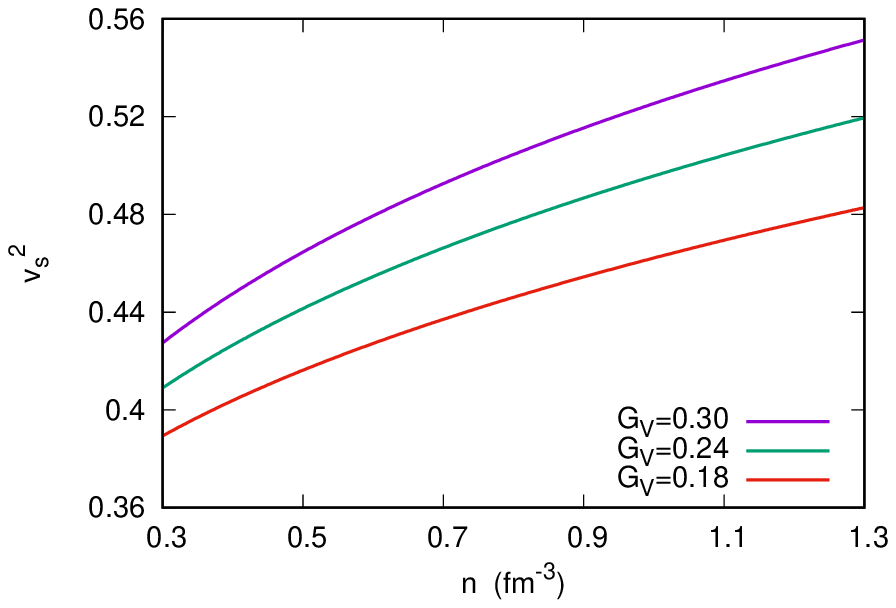} \\
\end{tabular}
\caption{(Color online) EoS (top) and the square of the speed of sound (bottom) for the three parametrizations discussed in the text. }  \label{Feos}
\end{centering}
\end{figure}

Another important quantity is the speed of sound of SQM, $v_s$, which provides information about the stiffness of the EoS related to the microscopic degrees of freedom of the model at hand. This in turn affects static and dynamic stellar observables \cite{Altiparmak:2022bke} such as tidal deformabilities and gravitational-wave signatures, respectively. Its square (for catalyzed matter) is defined as:
\begin{equation}
v_s^2 =  \frac{dp}{d\epsilon} .\label{sped} 
\end{equation}

We display in Fig.~\ref{Feos} the EoSs and the squared speed of sound for different values of $G_V$.

We can see that increasing the $G_V$ stiffens the EoS. If we choose $G_V$ = 0, therefore lower values of the bag would produce stiffer EoS, but the curves would have the same inclination, only displaced in the figure. The effect of the $G_V$ is more explicit in the speed of sound. If $G_V$ = 0, then all the EoSs would have the same $v_s^2$, independently of the bag value. However, as we increase $G_V$ we also increase the speed of sound. 

Moreover, in the absence of the Dirac sea~\cite{furnstahl1997vacuum}, both, the EoSs and the speed of sound are monotonically crescent with the density. If we simulate the Dirac sea contribution via a self-interaction in the vector field, we can have a decreasing speed of sound as pointed out in Ref.~\cite{lopesps2,Fran2024PRD}. However, as it softens the EoS we choose not to take into account the self-interaction.

\section{Configurations in relativistic hydrostatic equilibrum} \label{secIII}

In order to study the equilibrium configuration of a relativistic star, we begin by describing the background space-time using the following static, spherically symmetric line element given by
\begin{equation}\label{dsz_tov}
ds^{2}=-\mathrm{e}^{ \nu(r)} dt^{2} + \mathrm{e}^{ \lambda(r)} dr^{2} + r^{2}(d\theta^{2}+\sin^{2}{\theta}d\phi^{2})\,,
\end{equation}
\noindent where $t$, $r$, $\theta$ and $\phi$ are the set of Schwarzschild-like coordinates, and the metric potentials $\nu(r)$ and $\lambda(r)$ are functions of the radial coordinate $r$. This metric is used to compute the Einstein tensor $G_{\mu \nu}$. Then, we consider the energy-momentum tensor of a perfect fluid
\begin{equation}
T_{\mu \nu} = (\epsilon + p)u_{\mu}u_{\nu} + pg_{\mu \nu},
\end{equation}
where $u_{\mu}$ is the fluid 4-velocity and `$\epsilon$' and `$p$' are the energy density and pressure, respectively. Finally, we put the Einstein tensor $G_{\mu \nu}$ and this energy-momentum tensor into the Einstein field equations (in geometrical units) 
\begin{equation}
 G_{\mu \nu} =8\pi T_{\mu \nu},
\end{equation}
to obtain the Tolman-Oppenheimer-Volkoff (TOV) equations \cite{tov39,TOV}
\begin{eqnarray}
\frac{dm}{dr} &=& 4 \pi r^2 \epsilon,  \\
\frac{d\nu}{dr} &=&- \frac{2}{\epsilon} \frac{dp}{dr} 	\bigg(1 + \frac{p}{\epsilon}\bigg)^{-1},   \\
\frac{dp}{dr} &=& -\frac{\epsilon m}{r^2}\bigg(1 + \frac{p}{\epsilon}\bigg) \bigg(1 + \frac{4\pi p r^3}{m}\bigg)\bigg(1 - \frac{2m}{r}\bigg)^{-1},
\end{eqnarray}
\noindent where $m$ is the gravitational mass inside the radius $r$. These equations allow us obtain the stellar configuration in hydrostatic equilibrium and they will be used in this work for two main purposes. The first consists in determining the stellar masses and radii, while the second is aimed to study the behavior of physical quantities within the star, such as the pressure, energy density, speed-of-sound and metric radial profiles.

It is important to note that the metric function $\nu$ satisfies the following boundary condition
\begin{equation}\label{BoundaryConditionMetricFunction}
    \nu(r=R)= \ln \bigg( 1-\frac{2M}{R} \bigg),
\end{equation}
\noindent where $M$ and $R$ are the mass and radius for a given SQM star, respectively. With this condition, the metric function will match smoothly to the Schwarzschild metric outside the star according to the Birkhoff's theorem. The boundary conditions are $m(r=0) = 0$ and $p(r=R) = 0$.

\begin{figure}[ht]
  \begin{centering}
\begin{tabular}{c}
\includegraphics[width=0.333\textwidth,angle=270]{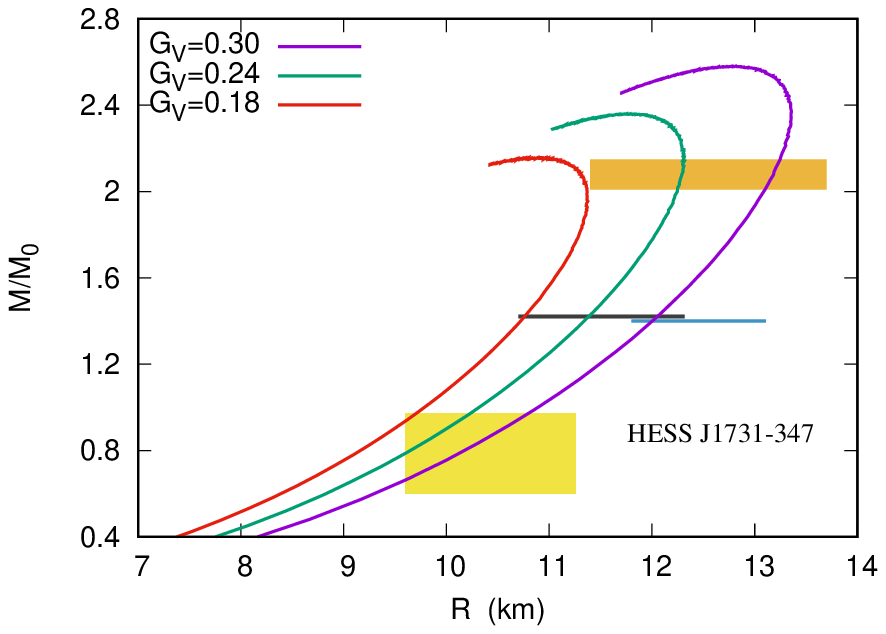} \\
\end{tabular}
\caption{Mass-radius diagram ($M_0$ is the Sun's mass) for SQM stars obtained by solving the TOV equations for the three parameterizations and constraints discussed in the main text.}  \label{Ftov}
\end{centering}
\end{figure}

We show in Fig.~\ref{Ftov} the resulting $M$-$R$ relation obtained by solving the TOV equations. We also discuss some constraints present in the literature. The first one is related to the radius of the 1.4 $M_\odot$ canonical neutron star.  Two NICER teams have pointed the stellar radius to a limit of $13.85 ~\mathrm{km}$ \cite{Riley:2019yda} and $14.26 ~\mathrm{km}$ \cite{Miller:2019cac}. These results were refined in Ref.~\cite{Miller2021}  to 11.80 km $<R_{1.4}< 13.10$ km. This constraint is represented by a narrow blue strip. Very close to this canonical mass, the mass and radius of the PSR J0437–4715 were recently constrained by one of the NICER teams in Ref.~\cite{J037}. The authors pointed out a mass of $M = 1.418 \pm 0.037M_\odot$ and a radius in the range $ R =11.36^{+0.95}_{-0.63}$ km, which presents a strong constraint for traditional hadronic stars. A black strip represents this astrophysics bound. Another important constraint is the well-established existence of very massive pulsars, with $M~>$ 2.0$M_\odot$. The more prominent example is the PSR J0740+6620 studied by two NICER teams, whose gravitational mass is 2.08 $\pm$ 0.07 $M_\odot$ with a radius lying in the range $R = 12.39^{+1.30}_{-0.98}$ km~\cite{Miller2021,Riley2021}. This constraint is presented as an orange-hatched area.  Finally, in the realm of very light objects, the proper existence of the so-called HESS J1731-347 supernova remnant present a puzzle once it has a unique mass of $M=0.77_{-0.17}^{+0.20}~M_\odot$ and a radius $R = 10.4 _{-0.78}^{+0.86}$ km~\cite{Doroshenko_2022}. This constraint
is presented as a yellow-hatched area.

We can see that while the HESS object presents a challenge for standard hadronic models, it can easily fit as a strange quark star. The three values of $G_V$ utilized in this work are able to fulfill such constraints. The same is true for the radius of the  PSR J0437–4715. On the other hand, the radius of the canonical star presented in Ref.~\cite{Miller2021}, can only be satisfied for $G_V$ = 0.30 fm$^2$. For lower values of the $G_V$, the radius of the 1.4$M_\odot$ is lower than 11.80 km. Finally, in the case of the PSR J0740+6620 with a mass of 2.08 $\pm$ 0.07 $M_\odot$, the $G_V$ = 0.18 cannot satisfy it. Although it predicts a maximum mass of 2.16 $M_\odot$,  all stars with $M~>$ 2.0 $M_\odot$ have a radius below 11.4 km. Ultimately, only $G_V$ = 0.30 fm$^2$ can simultaneously fulfill all these four constraints.

Another observable predicted by Einstein's general relativity is the gravitational redshift `$z$' which represents the fractional change between the observed (at infinity) and emitted (at the surface) wavelengths compared to the emitted wavelength. For a TOV star, it is given by~\cite{LopesEPL2021}:

\begin{equation}
 z =  \bigg ( 1 - \frac{2M}{R} \bigg )^{-1/2} - 1.   
\end{equation}

\begin{figure}[t]
\centering
\includegraphics[width=0.333\textwidth,angle=270]{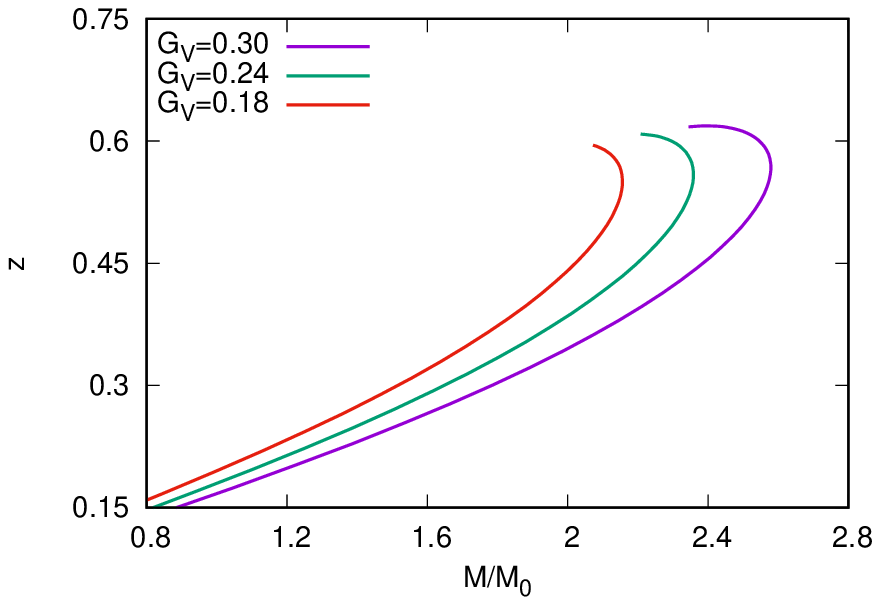}
\caption{The gravitational redshift `$z$' versus the SQM star's mass $M$ ($M_0$ is the Sun's mass) for different values of $G_V$.}\label{mxredshift}
\end{figure}

The results for the gravitational redshift of SQM stars from the vMIT model as a function of their masses are shown in Fig.~\ref{mxredshift} for different values of $G_V$. First, it is clear that the redshift enlarges as the total mass of the strange quark star grows. Second, it is easily seen that the redshift value strongly depends on $G_V$. Lower values of $G_V$ produce more compact stars, and consequently, larger values of $z$. Thus, by increasing the value of $G_V$, the redshift slightly decreases for SQM stars whose masses are less than 1$\,M_\odot$. As the mass of the stars grows, the differences in the redshift between each curve become more noticeable. When a strange quark star family approaches its maximum mass on each curve, they share almost the same value of $z$, although with different values of $M$.

\section{Radial oscillations} \label{secIV}

Radial oscillations are of significant interest in neutron star seismology because, for a star to exist, it must remain stable against radial perturbations. If the star becomes unstable, it could either explode or collapse, potentially forming a black hole. This underscores the importance of studying radial oscillations, and indeed, many studies exist on this subject, e.g. Refs. \cite{2018ApJ...860...12P,2019PhRvD.100b4035A,2020EPJC...80..726P}.

In order to analyze the radial oscillations of a self-bound star, we first need to establish its stellar configuration in hydrostatic equilibrium, as discussed in the previous section. Next, we apply an infinitesimal and radial disturbance (perturbation) to this configuration which keeps the star's spherical symmetry. These perturbations are introduced into the Einstein field equations, along with the conservation equations for energy, momentum, and baryon number. This process yields an Sturm-Liouville problem for the radial displacements \cite{APJ140:417:1964}.

Chronologically, this eigenvalue problem was originally derived by S. Chandrasekhar \cite{APJ140:417:1964} and later reformulated by Chanmugan \cite{APJ217:799:1977} in a format particularly well-suited for numerical applications, although with unintuitive boundary conditions. Two key quantities that describe these pulsations are the relative radial displacement, $\xi=\Delta r/r$, where $\Delta r$ represents the radial displacement of a matter element and $\Delta p / p$, where $\Delta p$ is the Lagrangian perturbation of pressure. Instead, in this work, we use the set of equations of Gondek et al. \cite{AA325:217:1997,PRD82:063006:2010} which is also a pair of differential equations which work nicely numerically due to the explicit use of $\Delta p$ as an unknown Lagrangian variable to be determined after applying intuitive boundary conditions. The equations read

\begin{eqnarray}\label{ecuacionparaXI}
\frac{d\xi}{dr} &=& -\frac{1}{r}\bigg(3\xi+\frac{\Delta p}{\gamma p}\bigg)-\frac{dp}{dr}\frac{\xi}{(p+\epsilon)},   \\
 \label{ecuacionparaP}
\frac{d\Delta p}{dr} &=& \xi \bigg\{\omega^{2}e^{\lambda-\nu}(p+\epsilon)r-4\frac{dp}{dr} \bigg \}  \nonumber \\  
&& +\xi\bigg \{\bigg(\frac{dp}{dr}\bigg)^{2}\frac{r}{(p+\epsilon)}-8\pi e^{\lambda}(p+\epsilon)pr \bigg \}   \nonumber \\ 
&& +\Delta p\bigg \{\frac{dp}{dr}\frac{1}{(p+\epsilon)} - 4\pi(p+\epsilon)r e^{\lambda}\bigg\},
\end{eqnarray}
\noindent where $\gamma$ is the relativistic adiabatic index, given by
\begin{equation}
    \gamma = \frac{(\epsilon + p)}{p} \bigg( \frac{dp}{d\epsilon} \bigg),
\end{equation}
$\omega$ represents the eigenfrequency, and the quantities $\xi \equiv \Delta r / r$ and $\Delta p$ are assumed to vary harmonically with time as $\mathrm{e}^{i\omega t}$.

 To solve equations (\ref{ecuacionparaXI}) and (\ref{ecuacionparaP}) one needs two boundary conditions. The first is the condition of regularity at the center ($r=0$), which can be written as \cite{AA260:250:1992,AA325:217:1997,AA344:117:1999}
\begin{equation}\label{DeltaP}
(\Delta p)_{\mathrm{center}}=-3(\xi \gamma p)_{\mathrm{center}}\,,
\end{equation}
\noindent where the eigenfunctions are normalized in order to have $\xi(0)=1$. The second boundary condition expresses the fact that the Lagrangian perturbation in the pressure at the stellar surface is zero
\begin{equation}
\label{PenSuperficie}
(\Delta p)_{\mathrm{surface}}=0,
\end{equation}
\noindent this condition arises from the requirement that the pressure reaches zero at the surface $p(r=R)=0$.
 
To numerically solve the oscillation equations, we proceed as follows. First, we choose a central pressure and integrate the Tolman-Oppenheimer-Volkoff (TOV) equations to obtain the stellar mass, radius, and also, the coefficients of the oscillation equations. Next, we apply the shooting method to solve the oscillation equations: starting with a trial value of $\omega^{2}$, we numerically integrate Eqs. (\ref{ecuacionparaXI}) and (\ref{ecuacionparaP}) using the initial values of $\xi(r=0)$ and $\Delta p(r=0)$ that satisfy the central boundary condition outlined above. The integration proceeds outward, with the aim of matching the boundary condition at the star's surface. After each integration, the trial value of $\omega^{2}$ is adjusted to improve the level of precision. The discrete values of $\omega^{2}$ that satisfy Eq. (\ref{PenSuperficie}) under these conditions are identified as the eigenfrequencies of the radial perturbations. For more details on the method see Ref. \cite{PRD82:063006:2010}. In this section, for reasons of coherence, we call $\nu_0$ the numerical value of the frequency of the fundamental radial mode.

\begin{figure}[t]
  \begin{centering}
\begin{tabular}{c}
\includegraphics[width=0.333\textwidth,angle=270]{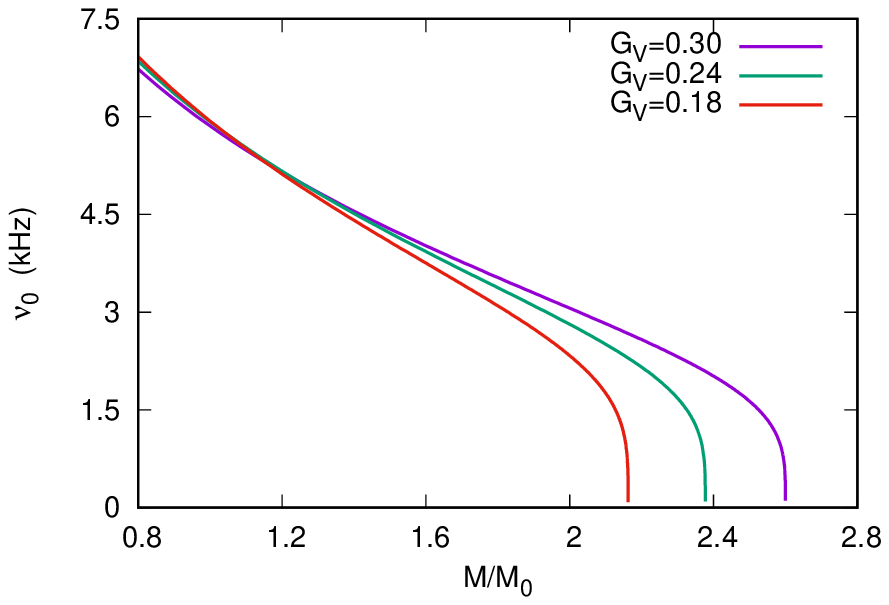} \\
\end{tabular}
\caption{The $f$-mode frequency ($\nu_0$) versus the gravitational mass $M$ ($M_0$ is the Sun's mass) for different values of $G_V$.}\label{radial}
\end{centering}
\end{figure}

In Fig. \ref{radial} we show our results for the radial $f$-mode frequency, $\nu_0$, as a function of the gravitational mass $M$ for different values of $G_V$. We can see that for stars with masses in the interval of approximately (1.2 - 2.0) $M_{\odot}$, the increase in the $G_V$ parameter produces an increase in the value of the fundamental radial mode. However, the increase in the $G_V$ parameter has the opposite effect below 1.2 $M_{\odot}$. We also observe that an increase in this parameter produces a shift to the right of all the massive stellar models, thus explaining why it is possible to obtain more massive and stable stars.
In fact, the increase in the parameter $G_V$ is related to the increase in mass, which can also be explained in terms of its stabilizing properties from a microscopic point of view. Therefore, we can conclude that increasing values of the $G_V$ parameter allow for more massive and dynamically stable massive self-bound stars obtained from the vMIT model.

We can also observe that the frequency reaches the zero value at the point where the models attain their maximum mass, as expected for one-phase stars. In contrast, if one allowed for a thin hadronic crust (not violating the Bodmer-Witten hypothesis), a discontinuous phase-transition-like phenomenon might occur near the stellar surface. However, since this crust is extremely thin, the whole stellar stability results for this radial $f$-mode will not be modified. Besides, our radial-oscillation calculation agrees with the static condition which only depends on the TOV solutions \cite{2018ApJ...860...12P}. In a future study, we plan to explore the presence of strong phase transitions inside neutron stars, but now using our vMIT model. This is because a recent work offers an alternative explanation for the low-mass ultracompact star HESS J1731-347, i.e. the authors of Ref. \cite{2024PhRvD.110d3026M} consider that this object has a phase transition potentially connected to quark deconfinement.

\section{Non-radial oscillations} \label{secV}

We now pass to study the non-radial perturbations of our self-bound stars obtained from the vMIT but only by focusing on the $f$-mode. It is worth to mention that we solve the full dynamical (time-dependent) problem within first-order perturbation theory without assuming any simplification such as the Cowling approximation.

Somewhat similar as in the radial case, we start now by applying polar non-radial perturbations to a non-rotating perfect fluid star giving a set of coupled equations \cite{1985ApJ...292...12D}. In this formalism, the perturbed metric tensor reads
\begin{eqnarray}
ds^2 & = & -e^{\nu}(1+r^{\ell}H_0Y^{\ell}_{m}e^{i\omega t})dt^2  - 2i\omega r^{\ell+1}H_1Y^{\ell}_me^{i\omega t}dtdr   \nonumber \\
&& + e^{\lambda}(1 - r^{\ell}H_0Y^{\ell}_{m}e^{i \omega t})dr^2 \nonumber \\
&& + r^2(1 - r^{\ell}KY^{\ell}_{m}e^{i \omega t})(d\theta^2 + \sin^2\theta d\phi^2),
\end{eqnarray}
where $H_0, H_1, K$ are metric perturbation functions and $\omega$ is the  angular frequency.

The polar perturbations in the position of the fluid elements are given  by the following Lagrangian displacements
\begin{eqnarray}
\xi^{r} &=& r^{\ell-1}e^{-\lambda/2}WY^{\ell}_{m}e^{i\omega t}, \\
\xi^{\theta} &=& -r^{\ell - 2}V\partial_{\theta}Y^{\ell}_{m}e^{i\omega t}, \\
\xi^{\phi} &=& -r^{\ell}(r \sin \theta)^{-2}V\partial_{\phi}Y^{\ell}_{m}e^{i\omega t} ,
\end{eqnarray}
where $W$ and $V$ are fluid perturbation functions. In the equations above $Y^{\ell}_{m}$ are the spherical harmonics. 

Using the previous quantities and putting them inside the perturbed Einstein equations, we can obtain the following set of first-order linear differential equations that govern the non-radial oscillations of the star \cite{1985ApJ...292...12D}:
\begin{eqnarray}
H_1' &=&  -r^{-1} [ \ell+1+ 2me^{\lambda}/r +4\pi   r^2 e^{\lambda}(p-\epsilon)] H_{1}   \nonumber \\ 
&& +  e^{\lambda}r^{-1}  \left[ H_0 + K - 16\pi(\epsilon+p)V \right] ,      \label{osc_eq_1}  \\
 K' &=&    r^{-1} H_0 + \frac{\ell(\ell+1)}{2r} H_1   - \left[ \frac{(\ell+1)}{r}  - \frac{\nu'}{2} \right] K  \nonumber \\
&&  - 8\pi(\epsilon+p) e^{\lambda/2}r^{-1} W \:,  \label{osc_eq_2} \\
 W' &=&  - (\ell+1)r^{-1} W   + r e^{\lambda/2} [ e^{-\nu/2} \gamma^{-1}p^{-1} X    \nonumber \\
&& - \ell(\ell+1)r^{-2} V + \tfrac{1}{2}H_0 + K ] \:,  \label{osc_eq_3}
\end{eqnarray}
and
\begin{eqnarray}
X' &=&  - \ell r^{-1} X + \frac{(\epsilon+p)e^{\nu/2}}{2}  \Bigg[ \left( r^{-1}+{\nu'}/{2} \right)H_{0}  \nonumber  \\
&& + \left(r\omega^2e^{-\nu} + \frac{\ell(\ell+1)}{2r} \right) H_1    + \left(\tfrac{3}{2}\nu' - r^{-1}\right) K   \nonumber  \\
&&    - \ell(\ell+1)r^{-2}\nu' V  -  2 r^{-1}   \Biggl( 4\pi(\epsilon+p)e^{\lambda/2} \nonumber  \\
&& + \omega^2e^{\lambda/2-\nu}  - \frac{r^2}{2}  (e^{-\lambda/2}r^{-2}\nu')' \Biggr) W \Bigg]  \:,
\label{osc_eq_4}
\end{eqnarray}
where the prime denotes a derivative with respect to the radial coordinate `$r$' and  `$\gamma$' is again the adiabatic index.
Also, in the equations above, the function $X$ is given by
\begin{eqnarray}
X =  \omega^2(\epsilon+p)e^{-\nu/2}V - \frac{p'}{r}e^{(\nu-\lambda)/2}W   \nonumber \\
  + \tfrac{1}{2}(\epsilon+p)e^{\nu/2}H_0 ,
\end{eqnarray}
and $H_{0}$ fulfills the algebraic relation 
\begin{eqnarray}
a_1  H_{0}= a_2  X -  a_3 H_{1}  + a_4 K,
\end{eqnarray}
with
\begin{eqnarray}
a_1 &=&  3m + \tfrac{1}{2}(l+2)(l-1)r + 4\pi r^{3}p  ,  \\
a_2 &=&  8\pi r^{3}e^{-\nu /2}   , \\
a_3 &=&  \tfrac{1}{2}l(l+1)(m+4\pi r^{3}p)-\omega^2 r^{3}e^{-(\lambda+\nu)}  ,  \\
a_4 &=&  \tfrac{1}{2}(l+2)(l-1)r - \omega^{2} r^{3}e^{-\nu}  \nonumber \\
    &&    -r^{-1}e^{\lambda}(m+4\pi r^{3}p)(3m - r + 4\pi r^{3}p)   .
\end{eqnarray}

Outside the star, i.e. the vacuum, \textcolor{blue}{$m= M$}, the perturbations equations on the fluid are null and the differential equations reduce to the very well-known Zerilli equations, which can be expressed as follows
\begin{equation}
\frac{d^{2}Z}{dr^{*2}}=[V_{Z}(r^{*})-\omega^{2}]Z,
\end{equation}
where $Z(r^{*})$ and $dZ(r^{*})/dr^{*}$  are related to the metric perturbations  $H_{0}(r)$ and $K(r)$ by the transformations given in Refs. \cite{2011ChPhB..20d0401L,1985ApJ...292...12D}.  We can also note the ``tortoise'' coordinate given by 
\begin{equation}
r^{*} = r + 2 M \ln (r/ (2M) -1), 
\end{equation}
and  the effective potential  $V_{Z}(r^{*})$ is given by
\begin{eqnarray}
V_{Z}(r^{*}) = \frac{(1-2M/r)}{r^{3}(nr + 3M)^{2}}f(r),
\end{eqnarray}
where
\begin{equation}
f(r)=[2n^{2}(n+1)r^{3} + 6n^{2}Mr^{2} + 18nM^{2}r + 18M^{3}]
\end{equation}
with $n= (l-1) (l+2) / 2$.

The system of Eqs. (\ref{osc_eq_1})$-$(\ref{osc_eq_4}) has four linearly independent solutions for given values of $l$ and $\omega$. In particular, we restrict ourselves to the $l = 2$ component, which dominates the emission of gravitational waves.  

The physical solution needs to verify the following appropriate boundary conditions:  
\begin{itemize}

\item  The perturbation functions: $H_1,K, W$ and $X$, must be finite everywhere, particularly at $r = 0$. To implement such a condition, it is necessary to use a power series expansion of the solution near the singular point $r=0$. The procedure is explained in detail in Refs. \cite{2011ChPhB..20d0401L,1985ApJ...292...12D}.

\item  The next boundary condition says that the Lagrangian perturbation in the pressure has to be zero at the surface of the star $r = R$. This implies that the function $X$ must vanish at $r = R$. 

\item  And finally, outside the star, the perturbed metric describes a combination of outgoing and ongoing gravitational waves.  The physical solution of the Zerilli equation is the one that describes purely outgoing gravitational radiation at $r=\infty$. 

\end{itemize}
Such boundary conditions cannot be verified by any value of $\omega$ and the frequencies that fulfill this requirement represent the quasinormal modes of the stellar model. Now we proceed to give our findings for these non-radial $f$-mode frequencies as well as their damping times $\tau$~\cite{Anderson1998}. The damping time, in the context of non-radial oscillations in compact stars, represents the characteristic timescale over which the energy of these oscillations dissipates. This dissipation is primarily driven by mechanisms such as gravitational wave emission, viscosity, and neutrino radiation. The damping time has a direct impact on the detectability of gravitational waves generated by oscillating compact stars. Short damping times result in a rapid decay of the gravitational wave signal, whereas longer damping times allow the signal to persist over extended periods, improving its likelihood of detection. In the era of advanced gravitational wave observatories like LIGO and Virgo, the damping time is a key parameter in modeling and interpreting signals from neutron stars and other compact objects. Within the theoretical framework, it is essential to emphasize that the damping time cannot be determined using the Cowling approximation~\cite{Prad2022}.

\begin{figure}[t]
  \begin{centering}
\begin{tabular}{c}
\includegraphics[width=0.333\textwidth,angle=270]{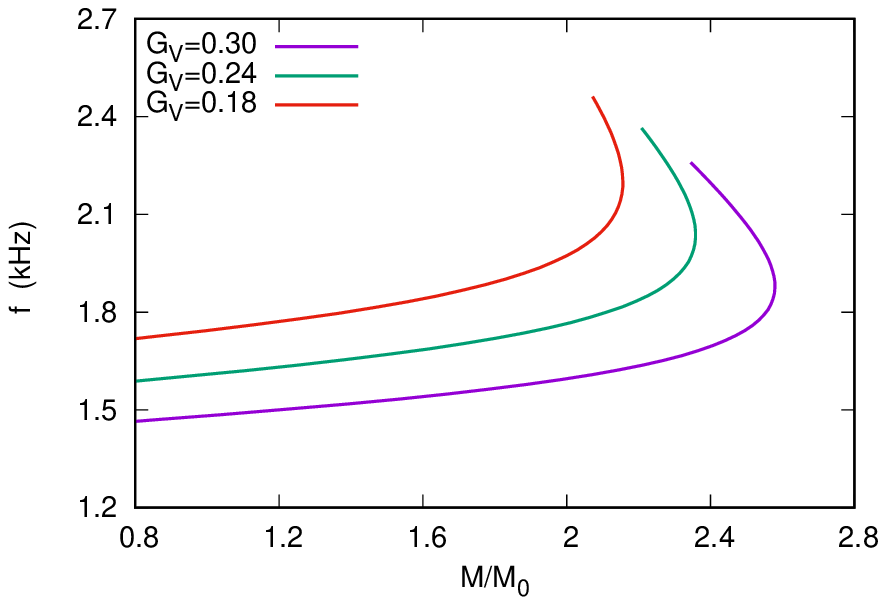} \\
\includegraphics[width=0.333\textwidth,,angle=270]{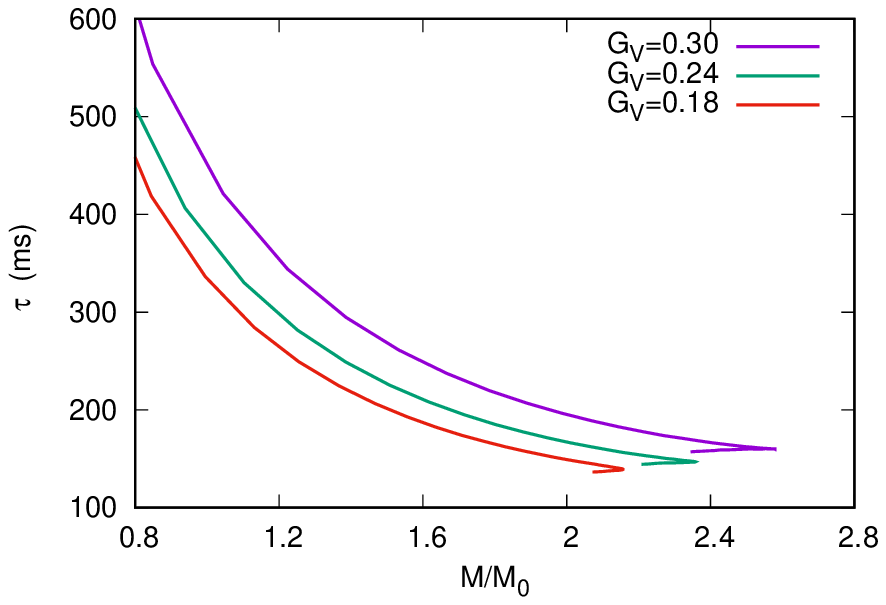} \\
\end{tabular}
\caption{Non-radial $f$-mode frequencies (top) and damping time (bottom) versus the mass $M$ for different values of $G_V$.}\label{fxm}
\end{centering}
\end{figure}

In the top panel of Fig. \ref{fxm} we show these $f$-mode frequencies. As it was explained in previous sections, our analysis considered three particular constant values of the parameter $G_V$ corresponding to the vMIT bag model. We can see that the increase in the $G_V$ parameter shifts the curves to the lower region of the figure. From this figure, it is also obvious that for masses below 2 $M_{\odot}$, the $f$-mode frequencies behave linearly with the gravitational mass. For SQM stars with masses in the range $2.0\, M_{\odot}-2.4\, M_{\odot}$, the $f$-mode curves upwards, i.e. the $f$-mode frequencies start to rise noticeably until the maximum mass of the stellar models is reached. It is important to mention that the NICER teams and the observations of the HESS object imposed strong restrictions and then the value $G_V$ = 0.30 fm$^2$ fulfills astronomical observations. Therefore, the constrained value of this parameter also imposes restrictions on the non-radial oscillations and gravitational frequencies. Then we can see that the frequency of the fundamental mode is restricted to (1.6 - 1.8) kHz for high mass stars and to (1.5 - 1.6) kHz for low mass stars.

Furthermore, we show the damping time $\tau$ for the $f$-mode in the lower panel of Fig. \ref{fxm}. One can see that the effect of the $G_V$ parameter has similar effects in the curves. For instance, near to the maximum mass of the stellar models, the $f$-mode curves downwards. The only difference is that this bending is more abrupt when compared to other work reported in the literature~\cite{Prad2022,lopescesar}. We also observe that an increase in the values of $G_V$ produces a systematic increase in the damping time, thus shifting the curves to the upper region of this figure.

\begin{figure}[t]
  \begin{centering}
\begin{tabular}{c}
\includegraphics[width=0.333\textwidth,angle=270]{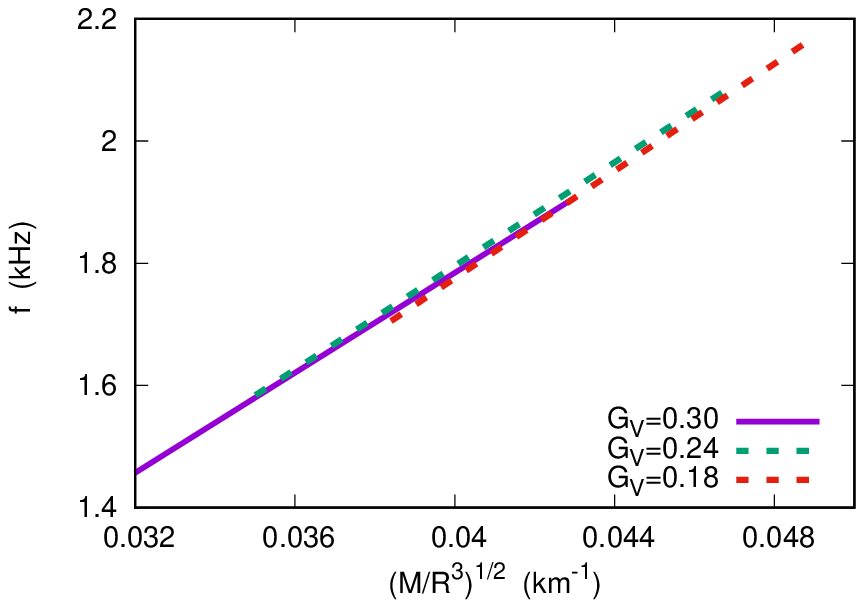} \\
\end{tabular}
\caption{ The frequency of the fundamental mode is plotted in the upper panel as a function of the square root of the average density for the different EoSs. The linear relation is satisfied for all three EoSs.}\label{dens}
\end{centering}
\end{figure}

To finish our analyses, we investigate the universal relation between the f-mode frequency, and the square root of the average density, $(M/R^3)^{1/2}$. It was pointed out in Ref.~\cite{Benhar2004} that in the Newtonian limit of the theory of stellar perturbations, the f-mode frequency scales as the
square root of the average density; and the $f$-mode can be fitted by the following linear expression:

\begin{equation}
 f = a + b\cdot(M/R^3)^{1/2},   \label{sqd}
\end{equation}
where $a$ is given in kHz and $b$ in km $\times$ kHz.  The results are displayed in Fig.~\ref{dens} for masses above 0.7 $M_\odot$.

We observe that the linear relation between $f$ and the square root of the average density is not only kept, but they present very similar coefficients, yet they cover different ranges of $f$. We obtain $a$ = $+0.142$, $+0.107$, and $+0.009$ kHz, $b$ = 41.1, 42.3, and 44.2 km $\times$ kHz for $G_V$ = 0.30, 0.24 and 0.18 fm$^2$ respectively. 

{Our mean value is $a = 0.086$ kHz and $b = 42.5$  km $\times$ kHz. We can compare these values with others present in the literature related to hadronic neutron stars. We compare our results with others found in the literature Tab.~\ref{T1}.}

\begin{center}
\begin{table}[ht]
\begin{center}
\begin{tabular}{c|cc}
\hline 
  Work & $a$ & $b$  \\
 \hline
 Anderson et al.~\cite{Anderson1998} & 0.220 & 47.5 \\
  Pradhan et al.~\cite{Prad2022} & 0.790  & 33.0   \\
 Benhar et al.~\cite{Benhar2004} & 0.535 & 36.2   \\
 Shirke et al.~\cite{Shirke2024PRD} &  0.630  & 33.5  \\
 Guha Roy et al.~\cite{Guha_2024APj} & 0.626 & 35.9 \\
 Chirenti et al.~\cite{Chirenti:2015dda} & 0.332 & 44.0 \\
 Flores et al.*~\cite{Cesar2017PRC} & -0.023 &  44.1 \\
\hline
This work* & 0.086 & 42.5    \\
\hline 
\end{tabular}
 
\caption{ Values of fitting coefficients for Eq.~\ref{sqd} from different works. Results market with * indicates strange stars.} 
\label{T1}
\end{center}
\end{table}
\end{center}

{ We notice that our value of angular coefficient $b$ is in agreement with those found in the literature. Furthermore, we can see that the coefficient $b$ seems very similar for hadron and quark stars. However, the constant $a$ for strange stars is much closer to zero than those related to hadron stars, probably due to the lack of the crust.}

\section{Conclusions} \label{conclusions}

As we discussed in our work, we have used the vector MIT model in order to verify the QM hypothesis, and as a consequence, it is possible to see that there exists little freedom for the bag constant $B$ once the $G_V$ parameter is selected, i.e all the results can be explained as a function of the $G_V$ parameter alone. It is important to mention that these parameter values are in agreement with recent astronomical observations.

Then, once we have selected the $G_V$ parameters, we explored the effect, of the vector field exchange term, on the mass and radius of strange quark stars. We have seen that the supernova remnant HESS J1731-347, and the PSR J0437-4715  are satisfied by the three $G_V$ parameters. But although we can always produce massive stars, with $M~>2.0M_\odot$, the radius of the PSR J0740-6620 is satisfied only by $G_V$ = 0.30 and 0.24 fm$^2$. We have found even more restrictions due to the radius of the canonical star, and finally, we concluded that $G_V$ = 0.30 fm$^2$ agrees with all constraints. Concerning the redshift $z$, we see that massive stars have a higher value of $z$. Moreover, for a fixed mass value, low values of $G_V$ produce a higher value of $z$. The redshift of the maximum massive star is almost the same for all values the $G_V$, although the masses are different.

In the sequence, we have studied the radial oscillations, because they are very important as a dynamical and strong test for the stability of a compact object. As a main conclusion we have seen that the increase in the $G_V$ parameter has a stabilizing property for stars in the range (1.2 - 2.0)M$\odot$ but for stars below 1.2 $M_\odot$ we have the opposite effect.

Finally, we studied the non-radial oscillations of strange quark stars. Recall that our results for the mass and radius of the astronomical observations have implicitly selected the $G_V$ parameter to the 0.30 fm$^2$ values. Then, as a main consequence, the gravitational wave frequency of the fundamental mode is restricted to (1.6 - 1.8) kHz for high mass stars and to (1.5 - 1.6) kHz for low mass stars. We also show that the linear relation between the $f$-mode and the square root of the average density expressed in Eq.~\ref{sqd} is satisfied for all EoS studied. { When compared with other values presented in the literature, we found that our value of $b$ is in agreement with them, while the value of $a$ is significantly lower seems much closer to zero for strange stars than for hadronic ones.}

\textbf{Acknowledgements:} L.L.L. was partially supported by CNPq Universal Grant No. 409029/2021-1. L.B.C. was partially supported by CNPq, Brazil under grants 308172/2023-0, FAPEMA and CAPES - Finance code 001. C. Flores 
acknowledges the financial support of the productivity program of the
Conselho Nacional de Desenvolvimento Científico e
Tecnológico (CNPq), with Project No. 304569/2022-4. J.C.J. is supported by Conselho Nacional de Desenvolvimento Cient\'ifico e Tecnológico (CNPq) with Grant No. 151390/2024-0.

\bibliography{abref}

\end{document}